\documentclass[fleqn,usenatbib]{rasti}

\usepackage{newtxtext,newtxmath}
\usepackage{pbox}
\usepackage{booktabs}
\usepackage{orcidlink}

\usepackage[T1]{fontenc}

\DeclareRobustCommand{\VAN}[3]{#2}
\let\VANthebibliography\thebibliography
\def\thebibliography{\DeclareRobustCommand{\VAN}[3]{##3}\VANthebibliography}

\usepackage{graphicx}	
\usepackage{amsmath}	
\usepackage{color,soul}
\usepackage{hyperref}
\usepackage{algorithm}
\usepackage{algpseudocode}

\usepackage{tikz}
\usetikzlibrary{shapes.geometric, arrows, decorations.pathreplacing, calc, 3d}

\tikzstyle{startstop} = [rectangle, rounded corners, minimum width=3cm, minimum height=1cm,text centered, draw=black, fill=red!30]
\tikzstyle{process} = [rectangle, minimum width=3cm, minimum height=1cm, text centered, draw=black, fill=blue!30]
\tikzstyle{decision} = [diamond, minimum width=3cm, minimum height=1cm, text centered, draw=black, fill=green!30]
\tikzstyle{arrow} = [thick,->,>=stealth]

\newcommand{\pymgal}{\textsc{PyMGal}}

\title[\textsc{PyMGal}]{\pymgal: A Python Package for Generating Optical Mock Observations from Hydrodynamical Simulations}

\author[Janulewicz \& Cui]{
Patrick Janulewicz \orcidlink{0009-0004-9465-428X}$^{1,2,3,4}$\thanks{E-mail: patrick.janulewicz@mail.mcgill.ca}, 
Weiguang Cui \orcidlink{0000-0002-2113-4863}$^{5,6,7}$\thanks{E-mail: weiguang.cui@uam.es}
\\
$^{1}$Department of Physics, McGill University, Montreal H3A 2T8, Canada  \\
$^{2}$Trottier Space Institute, McGill University, Montreal H3A 2A7, Canada  \\
$^{3}$Ciela Institute, Montreal  H2V 0B3, Canada \\
$^{4}$ Mila - Quebec Artificial Intelligence Institute, Montreal H2S 3H1, Canada \\
$^{5}$ Departamento de Física Teórica, Módulo 15, Facultad de Ciencias, Universidad Autónoma de Madrid, 28049 Madrid, Spain \\
$^{6}$ Centro de Investigación Avanzada en Física Fundamental (CIAFF), Facultad de Ciencias, Universidad Autónoma de Madrid, 28049 
Madrid, Spain \\
${^7}$ Institute for Astronomy, University of Edinburgh, Royal Observatory, Blackford Hill, Edinburgh EH9 3HJ, UK \\
}

\date{Accepted XXX. Received YYY; in original form ZZZ}

\pubyear{\the\year{}}

\begin{document}
\label{firstpage}
\pagerange{\pageref{firstpage}--\pageref{lastpage}}
\maketitle

\begin{abstract}

We introduce \pymgal, a Python package for generating optical mock observations of galaxies from hydrodynamical simulations. \pymgal\ reads the properties of stellar particles from these simulations and generates spectral energy distributions (SEDs) based on a variety of stellar population models that can be customised to fit the user's choice of applications. Given these SEDs, the program can calculate the brightness of particles in different output units for hundreds of unique filters. These quantities can then be projected to a 2D plane mimicking a telescope observation. The software is compatible with different snapshot formats and allows a flexible selection of models, filters, output units, axes of projection, angular resolutions, fields of view, and more. It also supports additional features including dust attenuation, particle smoothing, and the option to output spectral data cubes and maps of mass, age, and metallicity. These synthetic observations can be used to directly compare the simulated objects to reality in order to model galaxy evolution, study different theoretical models, and investigate different observational effects. This package allows the user to perform fast and consistent comparisons between simulation and observation, leading to a better and more precise understanding of what we see in our Universe.
\end{abstract}




\section{Introduction}

Hydrodynamical simulations provide key insights to enhance our understanding of the Universe. By solving the equations of gravity and hydrodynamics within a discretised box, they are used to trace the distribution of matter and the properties of astrophysical objects from early times to the modern era. These simulations have evolved significantly since they were first introduced, with many different codes and formats available. These include \textsc{GADGET} \citep{gadget} and its subsequent versions and variants, such as \textsc{GIZMO} \citep{gizmo}, \textsc{Arepo} \citep{Springel2010}, and \textsc{GADGET-4} \citep{Springel2021}. Besides the \textsc{GADGET} family, many other numerical simulation programs, such as \textsc{RAMSES} \citep{ramses}, \textsc{SWIFT} \citep{swift}, \textsc{ENZO} \citep{enzo}, and \textsc{Pkdgrav3} \citep{Potter2017}, offer different features, including baryonic physics and hydrodynamics solvers, or are designed with different scientific goals in mind. These simulations generally track the matter in the universe, such as dark matter and baryons, through different particles. As these codes follow the evolution of a system, this particle information, such as position and velocity, is saved in discrete time steps. These steps, commonly referred to as snapshots, contain the relevant data describing the simulated universe at that point in time. As such, the matter distribution and dynamical information of the simulation can be mapped out through the sample particles. Different simulation codes may have different formatting for the snapshot data, with each one potentially requiring unique handling when being read.

Modern hydrodynamical simulation projects, such as Illustris \citep{illustris}, EAGLE \citep{eagle}, IllustrisTNG \citep{tng}, SIMBA \citep{simba} and many others, also provide the baryonic information through the gas and star particles. Due to the computational demand of these simulations, these particles represent populations of astronomical objects rather than being individually resolved. Each stellar particle in a simulation, for instance, represents a group of coeval stars known as a simple stellar population (SSP). While the information provided for stellar particles includes physical quantities such as position, mass, age, and metallicity, it generally does not contain the brightness of a given particle, which is directly measured by the observer. This creates a barrier between simulation and observation. To overcome this problem, one often needs to infer the spectrum based on the different physical properties of the stellar particles. The light emitted from these particles at different wavelengths is described by the spectral energy distribution (SED). Stellar population synthesis (SPS), which provides a framework for modelling the SEDs of stellar populations based on their physical characteristics, can then be leveraged to infer the observable properties of these simulated stellar particles. Modelling stellar populations, however, is a challenging task, as many unknowns exist within the field. Different design choices must be made to handle these unknowns, including the initial mass function (IMF), the star formation history (SFH), and the treatment of stellar evolution models.

Many different solutions have been proposed to visualise the contents of these simulations, typically by using existing SPS libraries to infer SEDs, incorporating relevant physical processes, and projecting the resulting data. The most complete approaches, which include radiative transfer processes, such as \textsc{powderday} \citep{Narayanan2021} and \textsc{GRASIL-3D} \citep{Dominguez2014}, are very time-consuming. Therefore, other packages, such as \textsc{FORECAST} \citep{Fortuni2023} and \textsc{Synthesizer} \citep{Lovell2024} have been developed. These packages provide a link that allows consistent comparisons between simulation and observation. In this work, we present a fast and flexible approach for creating these synthetic observations that facilitates comparison between different SPS models and addresses the many uncertainties associated with modelling stellar populations.

A simple solution to perform these comparisons is proposed by EzGal \citep{ezgal}, which provides tools to calculate magnitudes and other observables across a range of stellar population models from various works. Included among these is the model set introduced in \citeauthor{bc03} (\citeyear{bc03}; hereafter BC03), which is one of the most widely used model sets in the literature. Another is the set from \citeauthor{m05} (\citeyear{m05}; hereafter M05), which thoroughly treats thermally pulsating asymptotic giant branch (TP-AGB) stars due to their importance in the infrared emission of young stellar populations. EzGal also features a subsequent version of the BC03 models described in works such as \citeauthor{cb07} (\citeyear{cb07}; hereafter CB07), which updates the BC03 models to treat these TP-AGB stars. The $\alpha$-enhanced BaSTI models by \citeauthor{basti} (\citeyear{basti}; hereafter P09) and the customisable FSPS models from \citeauthor{conroy_gunn_white_2009} (\citeyear{conroy_gunn_white_2009}; hereafter C09) and \cite{conroy_gunn_2010} are included as well. Finally, the PEGASE.2 set from \citeauthor{pegase2} (\citeyear{pegase2}; hereafter P2), was also added to this list after publication. In addition to the sets mentioned above, EzGal also supports custom models, enabling users to tailor them to specific scientific needs while ensuring adaptability to future advancements in the field. These models can be used in combination with an initial mass function such as \cite{salp_imf}, \cite{kroupa_imf}, or \cite{chab_imf} to model SEDs of different stellar populations. EzGal also provides a framework for constructing composite stellar population (CSP) models from these SSP models, which may involve more sophisticated star formation histories. EzGal is often used in observation to fit the SED of a galaxy to derive its physical quantities, such as mass, metallicity and SFH.

However, from the simulator's point of view, the problem is different. The physical quantities for cosmological simulations are already known, but one would wish to see what the simulated galaxies look like through different telescopes. Therefore, SPS can be leveraged to model SEDs for simulated stellar particles given some assumed model by using EzGal reversely. By combining this with the flexible approach described above, we are able to predict SEDs in cosmological simulations for any arbitrary choice of stellar population model. Using these SEDs and some assumed filter response curves, we can calculate the magnitudes of simulated stellar particles as would be seen through different photometric bands. With magnitudes calculated and positions provided, the information can be combined in order to visualise the inside of the simulation box. 

In this paper, we present \pymgal, a Python program to generate mock observations of hydrodynamical simulations in the optical regime. \pymgal\ performs forward modelling by leveraging SPS to convert the physical properties of simulated stellar particles into observable quantities, enabling direct comparison between simulations and real astronomical observations. These projections can be made in various output units along any arbitrary axis, with over one hundred filters included in the package. It is compatible with simulation formats such as \textsc{GADGET}, \textsc{GIZMO}, \textsc{AREPO}, and more, and it can also be extended to other formats if needed.

One of the key strengths of \pymgal\ is its ability to incorporate any choice of stellar population model, regardless of its original library, underlying assumptions, or parameter choices. This ensures that users are not restricted to any predefined set but can instead implement the most suitable framework for their research, thus facilitating comparisons across different models and enabling tailored selection based on specific scientific goals. Another advantage is the flexibility in the data products that it can output. In addition to supporting a wide range of output units for optical maps, the package allows users to generate maps of mass, age, and metallicity that align with these optical observations. It also features the ability to generate spectral data cubes for these projections, allowing for the detailed study of the object's intrinsic spectrum along different viewing angles. Moreover, \pymgal\ is optimised for speed and efficiency. It has a lightweight design that allows for fast processing of large datasets without compromising the flexibility and range of outputs.

The paper is organised as follows. In Section \ref{sec:modelling}, we describe the approach taken to model stellar populations and generate SEDs for simulated particles. In Section \ref{sec:filters_and_mags}, we explain the methods used to calculate magnitudes in a photometric band given a filter response curve. In Section \ref{sec:projections}, we describe the techniques used to generate projections given these particles and their computed magnitudes. We also describe optional modifiable parameters in this section such as smoothing, the addition of spectra, mass, age, and metallicity maps, and more. In Section \ref{sec:validation}, we demonstrate the reliability of \pymgal's outputs by comparing to an established benchmark. Finally, in Section \ref{sec:conclusions}, we conclude and discuss the applications of this software to modern problems.

\section{Modelling Stellar Populations}\label{sec:modelling}

\subsection{Reading data and generating SEDs}\label{sec:seds}

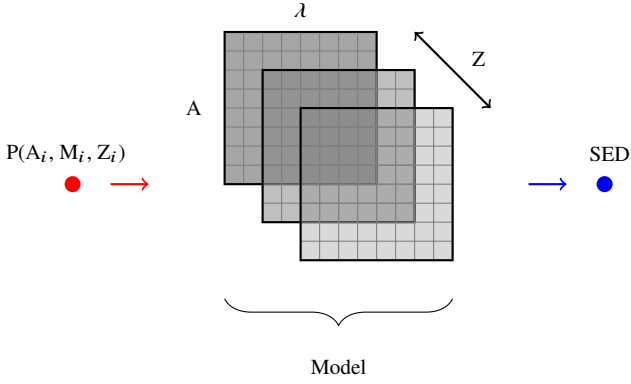
\begin{figure}
    \centering
    \begin{tikzpicture}

    \draw[fill=gray, opacity=0.7] (0,0) rectangle (2,2); 
    \draw[step=0.25, black!50] (0,0) grid (2,2); 
    \draw[draw=black, thick] (0,0) rectangle (2,2); 

    \draw[fill=gray, opacity=0.5] (0.5,-0.5) rectangle (2.5,1.5); 
    \draw[step=0.25, black!50] (0.5,-0.5) grid (2.5,1.5); 
    \draw[draw=black, thick] (0.5,-0.5) rectangle (2.5,1.5); 

    \draw[fill=gray, opacity=0.3] (1,-1) rectangle (3,1); 
    \draw[step=0.25, black!50] (1,-1) grid (3,1); 
    \draw[draw=black, thick] (1,-1) rectangle (3,1); 

    \draw[thick,<->] (2.5,2) -- (3.5,1) node[midway,right=0.15cm, yshift=0.15cm] {Z}; 

    \node[anchor=east] at (-0.2,1) {A};

    \node[anchor=south] at (1,2.1) {$\lambda$};

    \fill[red] (-2, 0) circle (3pt); 
    \node[anchor=east] at (-1.20, 0.4) {P(A$_i$, M$_i$, Z$_i$)}; 

    \draw[thick,->,red] (-1.5, 0) -- (-1, 0); 

    \fill[blue] (5, 0) circle (3pt); 
    \node[anchor=west] at (4.7, 0.4) {SED}; 

    \draw[thick,->,blue] (4, 0) -- (4.5, 0); 

    \draw [decorate,decoration={brace,amplitude=10pt,mirror,raise=4ex}]
     (0,-1) -- (3,-1) node[midway,yshift=-5em]{Model};

    \end{tikzpicture}
    \caption{Visualisation of how SEDs can be inferred given the mass, age, and metallicity of a particle. Each square represents a 2D array indexed by wavelength $\lambda$ and age A. The particle P with metallicity Z$_i$ is assigned to the closest metallicity value. The shape of the SED is then inferred given its age A$_i$, and it is later scaled by its mass M$_i$.}
    \label{fig:assigning_seds}
\end{figure}

The first data product required by the software is the SPS library. Though the appearance of this distribution is dependent upon the choice of model, all share a common structure. Each one can be represented by a two-dimensional array containing ages on one axis and wavelengths or frequencies on the other. The value at any given age-wavelength or age-frequency pair will describe the energy output of the object being studied. However, these SEDs vary as a function of metallicity as well. To address this, the metallicity range can be divided into discrete levels, with each level being assigned its own file containing a two-dimensional SED array. Particles within the simulation, on the other hand, may have a more continuous range of such values. To resolve this, the SEDs will be interpolated, allowing the software to assign SEDs to particles given their age and metallicity. We illustrate this process in Figure \ref{fig:assigning_seds}.

\pymgal\ can then begin to assign SEDs to particles within the simulation. To do this, data must be read from the snapshot file, which provides the coordinates and physical properties of particles. The program can read the data given a centre and radius representing a region of the box. A volume is then computed given this information, and all stellar particles within it are then handled by the software. Any particles falling outside the specified region will not be considered. This can be modified by the user to encompass any arbitrary region, including the entire simulation box. The age and metallicities of the particles are of particular interest for the task of assigning SEDs, and the resulting SEDs are scaled based on the mass of the particle. To perform this scaling, the SED is first divided by the normalised mass provided in the model file and then multiplied by the mass of the simulated stellar particle. This ensures proper treatment of the mass loss due to evolution. \pymgal\ also reads the cosmology of the simulation, which is required to calculate different values including age, redshift, and distance. Given this information, particle SEDs can be inferred, and the program can move on to the next step.

When using the software, the user should ensure that the selected SPS library encompasses the full range of metallicities present in the simulation to maintain consistency in the assigned SEDs. If the metallicity range of the models is insufficient, values outside this range will be mapped to the nearest boundary and may not capture the true nature of the particles. In such cases, an SPS library with extended metallicity values may be used to better match the properties of the simulated stellar particles.

\subsection{Model selection}
\label{sec:models} 

\pymgal\ must be provided with a model on which it bases its SEDs. The software supports different formats for these model files.

To begin, the software is compatible with all EzGal SSP and CSP models, which are distributed in EzGal's \texttt{.model} format and are readily convertible to the more standard Flexible Image Transport System \texttt{.fits} format. It comes pre-installed with many of these files, though more can be added if needed. The SSP model files are defined by the model type, the IMF, and the metallicity. Model types include BC03, M05, CB07, P09, C09, and P2, while choices of IMF include Salpeter, Kroupa, and Chabrier. Each model file will contain the SED's evolution at various ages for a given metallicity. \pymgal\ is also compatible with the binary \texttt{.ised} format in which the BC03 and CB07 models are distributed. Finally, to maximise flexibility, \pymgal\ can also read files in the ASCII \texttt{.txt} format. In theory, these options allow the user to implement any arbitrary model with any arbitrary set of parameters. The user may define such a custom file and add it in the relevant directory.

This file, or set of files, will then be read by the program and used in interpolation. A list of EzGal models also supported by \pymgal\ can be found in Table \ref{tab:models}. Users may also add additional models to match their research goals.

The effect of nebular emission may play an important role in broadband imaging data. Users wishing to incorporate this effect are encouraged to use SPS models designed to replicate the impact of nebular emission. For instance, software such as FSPS and works such as \cite{gutkin_emission_line} provide a framework for modelling emission lines in stellar SEDs. \pymgal\ allows users to input such models, ensuring that these effects are taken into account in the simulated observations.

\begin{table}
\centering
\begin{tabular}{|l|l|l|c|c|}
\toprule
\multicolumn{1}{|c|}{\textbf{Category}} & \multicolumn{1}{c|}{\textbf{IMF}} & \multicolumn{1}{c|}{\textbf{$\text{Z}/\text{Z}_{\odot}$}} &  
\textbf{No.\ Z} &  \textbf{No.\ ages}  \\
\midrule
BC03             & \pbox{1cm}{Chabrier Salpeter} & 0.005 - 2.5 &   6 & 221  \\
\midrule 
M05                          &  \pbox{1cm}{Kroupa Salpeter} & 0.05 - 3.5  & 5  &   68  \\
\midrule
CB07                          &  \pbox{1cm}{Chabrier Salpeter} & 0.005 - 2.5 & 6  & 221  \\
\midrule
C09                           &  \pbox{1cm}{Chabrier Salpeter Kroupa}  & 0.01 - 1.5   & 22        & 189 \\ 
\midrule
P09           & Kroupa   & 0.005 - 2.5  & 10         & 56  \\
\midrule
P2                            &  Salpeter   & 0.005 - 5         & 
7 & 69 \\
\bottomrule
\end{tabular}
\caption{EzGal-formatted SPS libraries that are also supported by \pymgal. The first column indicates the work from which the model was obtained, the second column specifies the available initial mass functions, and the third column indicates the range of metallicities covered by the library. The fourth and fifth columns indicate the number of metallicities and ages, respectively, featured by the library. }
\label{tab:models}
\end{table}

\subsection{Dust models}
\label{sec:dust}

The effect of dust attenuation can optionally be included. Though some SPS software provides the option to apply dust attenuation immediately upon calculating the SEDs, this may not always be the case. We therefore provide two simple models. 

A typical example of dust attenuation is described by \cite{charlot_fall}. The impact of dust as a function of wavelength $\Gamma(\lambda)$ can be described mathematically using Equation \ref{eq:charlot_fall}.

\begin{equation}
    \Gamma(\lambda) = e^{-\tau(t)(\lambda/ 5500\text{Å} )^{-n}} \label{eq:charlot_fall}
\end{equation}

In the above expression, we have $\tau(t) = 1.0$ for $t \leq t_{\text{break}}$ and $\tau(t) = 0.5$ for $t > t_{\text{break}}$ and a power law index $n$ that is typically assigned a value of $0.7$. Moreover, $t_{\text{break}}$ is a cutoff time that differentiates younger stars from older ones, often assigned a value of $10^7$ years. For stars younger than the cutoff time, contributions from both the birth cloud and the interstellar medium are considered, while stars older than this time model only contributions from the interstellar medium. 

Another common example is the dust function from \cite{calzetti}, the effect of which is described in Equation \ref{eq:calzetti}, where $k'(\lambda)$ is the starburst reddening curve and $E_s(B-V)$ is the colour excess of the stellar continuum. 

\begin{equation}
    \Gamma(\lambda) = 10^{-0.4 E_s(B-V)k'(\lambda)} \label{eq:calzetti}
\end{equation}

While \pymgal\ provides built-in support for these two functions, others may also be added. Users may define and implement a custom dust function that suits the specific needs of their research applications in the program.

In addition to the above models, \pymgal\ may also be used to calculate the effect of dust along the line of sight. Rather than employing computationally expensive radiative transfer mechanisms, we adopt a fast approach by modelling dust based on the gas particles along each stellar particle’s line of sight. This method is designed to ensure computational efficiency while providing a fast yet adaptable treatment of line-of-sight dust attenuation. 

If this option is selected, the software will read the properties and positions of gas particles once they are projected using the process described in Section \ref{sec:proj_overview}. For each projection, both stellar and gas particles are binned to a 2D histogram, and the bin assignments are stored by the program. The software identifies gas particles that share the same histogram bin and are positioned between a given stellar particle and the observer, ignoring those located behind it.

With the relevant gas particles identified, \pymgal\ then computes the effect they will have on their associated stellar particle. This effect will dim the brightness of a particle by a factor of $e^{-\tau}$ for an optical depth $\tau = \int \kappa \rho ds$, where $\kappa$ is the opacity, $\rho$ is the density along the line of sight, and $s$ is the path from the particle to the observer. The exact treatment of this optical depth will then depend on the user's choice of assumptions. In the simplifying assumption of constant opacity and density along the path, the optical depth may be written  $\tau = \kappa \frac{m_g}{A}$, where $m_g$ is the total gas mass along the line of sight within a pixel of area $A$. The effect of dust in this case can be written in the form of Equation \ref{eq:los_dust} assuming a wavelength-dependent opacity.

\begin{equation}
    \Gamma(\lambda) = e^{-\kappa(\lambda) \frac{m_g}{A}} \label{eq:los_dust}
\end{equation}

The user's preferred treatment for the opacity will help determine the value of the optical depth. This opacity may be specified by the user or inferred from the properties of the gas particles. When inferring this value based on gas properties, there exist several different sources of opacity that may dominate in different environments. For instance, in highly ionised gas, electron scattering is a dominant opacity source, which we denote $\kappa_e$. Free-free, bound-free, and bound-bound electron transitions become important in regimes where the material is partially ionised, and their contributions can be approximated using Kramers opacity $\kappa_K$. Moreover, in solar atmospheres, opacity due to the negative hydrogen ion $\kappa_{H^{-}}$ dominates. Finally, opacity from dust $\kappa_d$ also has an important impact in regions such as the interstellar medium where dust grains are abundant. The user may choose to consider contributions from one or several of these sources. Mathematical descriptions of these opacity sources in units of cm\textsuperscript{2} g\textsuperscript{-1} are shown in Equations \ref{eq:kappa_e} to \ref{eq:kappa_d}. The first three sources are defined for a gas of density $\rho$, temperature $T$, hydrogen mass fraction $X$, and metallicity $Z$ \citep{stellar_interiors_textbook}. For the dust opacity, we provide a flexible power law scaling given an exponent $\gamma$ and a reference opacity $\kappa_0$ at a reference wavelength $\lambda_0$. A typical exponent for the interstellar medium is given by $\gamma \approx 1.7$ with $\kappa_0 = 0.01 $ cm\textsuperscript{2} g\textsuperscript{-1} based on dust mass at $\lambda = 250$ $\mu$m \citep{draine_opacity_2006, hildebrand_opacity_1983}, though the user may modify these to best match the conditions of the simulation. As \pymgal\ continues to be developed, we expect to expand upon the treatment of opacities and extend its capabilities to even broader ranges of astrophysical conditions.

\begin{align}
    \kappa_e &\approx  0.2 (1 + X) \label{eq:kappa_e} \\
    \kappa_K &\approx 4 \times  10^{25} Z (1 + X)  \rho T^{-3.5} \label{eq:kappa_K}\\
    \kappa_{H^{-}} &\approx 2.5 \times 10^{-31} \left(\frac{Z}{0.02}\right) \rho^{1/2} T^9 \label{eq:kappa_h}\\
    \kappa_d & \approx \kappa_0  \left(\frac{\lambda}{\lambda_0}\right)^\gamma \label{eq:kappa_d} 
\end{align}

Given these preferences, one opacity value is computed for each gas particle between the associated stellar particle and the observer, and an average opacity is determined. The total mass of these gas particles $m_g$, as well as the pixel area $A$, are also computed. With these values, the optical depth for a stellar particle can then be estimated and used to dim the brightness of the stellar particles to provide a more realistic rendition of the synthetic observation.

\section{Filters and magnitudes}
\label{sec:filters_and_mags}

\subsection{Calculating magnitudes}
\label{sec:magnitudes}

With the SEDs computed and the optional dust attenuation function applied, the next step is to calculate magnitudes. The software uses the approach from EzGal, which calculates magnitudes in the observer's frame as a function of redshift $z$ and formation redshift $z_f$. The age of a galaxy can then be denoted $t(z, z_f) = T_U(z) - T_U(z_f)$ where $T_U(z)$ is the age of the universe as a function of redshift given the cosmology of the simulation, and the redshifted SED at this age can then be written $F_\nu[\nu(1+z), t(z, z_f)]$. The magnitude can then be defined as the projection of the redshifted SED through a filter response curve $R(\nu)$, compared to a zero magnitude AB source. This calculation is illustrated in Equation \ref{eq:mag_calc}. In this equation, the SED has units of erg s\textsuperscript{-1} cm\textsuperscript{-2} Hz \textsuperscript{-1} for an object at a distance of 10 pc.

\begin{align}
     &M_{\text{AB}}[z, t(z, z_f )] = \label{eq:mag_calc}\\
     &-2.5\log\left[\frac{\int_{-\infty}^{\infty} \nu^{-1} (z+1)F_{\nu}[\nu(1+z), t(z, z_f)] R({\nu}) d\nu}{\int_{-\infty}^{\infty} \nu^{-1} R(\nu) d\nu}\right]  - 48.60 \notag
\end{align}

The rest-frame magnitude uses the same approach, but calculates the magnitude as $M_{\text{AB}}[0, t(z, z_f )]$.

\pymgal\ supports conversions to other units and magnitude systems. It can convert to Vega or solar magnitude, as well as different units of flux, flux density, and luminosity. More details on output units and conversions can be found in Section \ref{sec:outputs}.

\begin{figure}
    \includegraphics[width=\columnwidth]{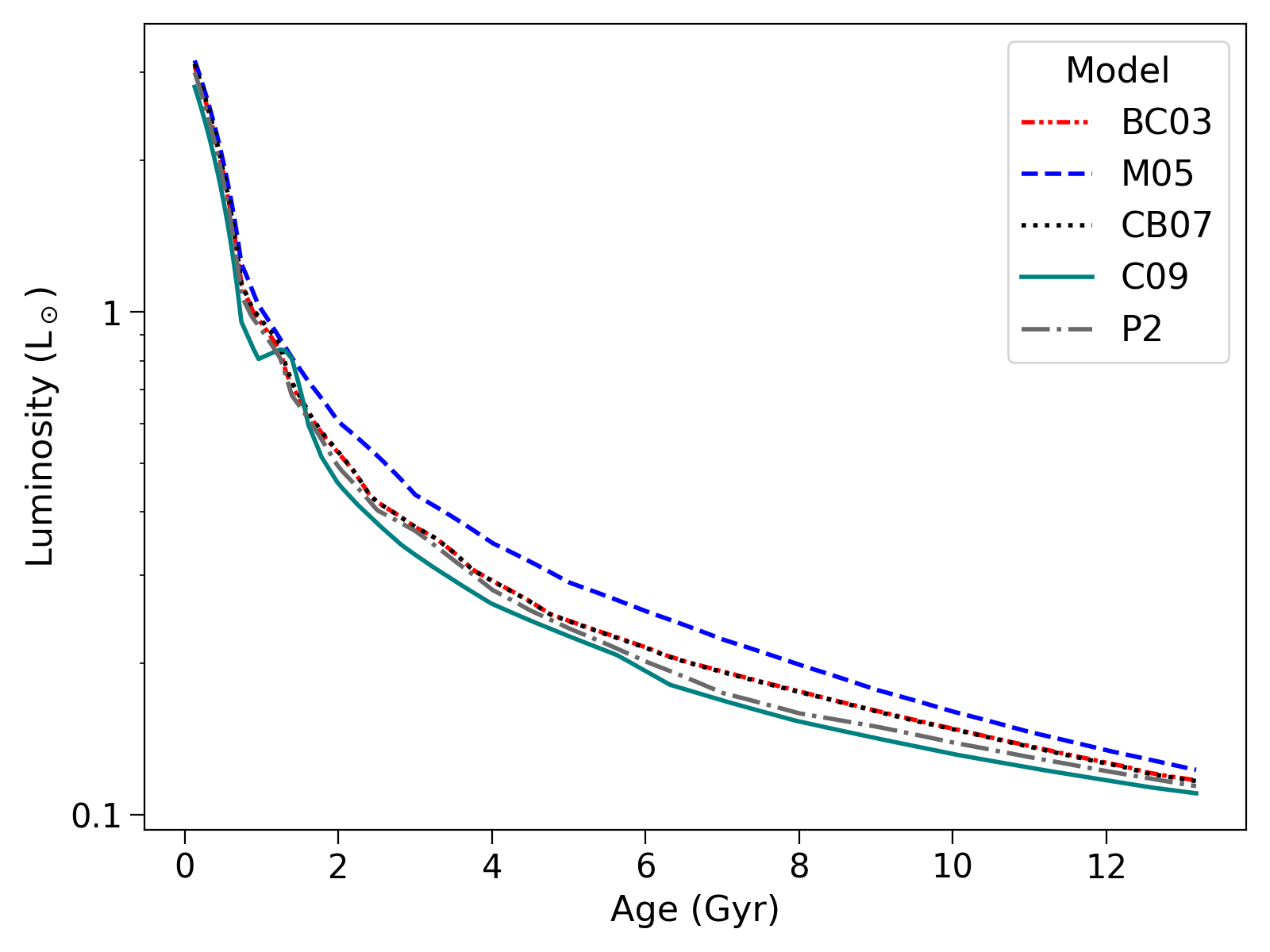}
    \caption{The luminosity evolution of a 1 $M$\textsubscript{$\sun$} stellar particle across various models, all assuming a Salpeter IMF and solar metallicity. The models exhibit minor variations but remain in reasonable agreement.} 
    \label{fig:model_comp}
\end{figure}

We provide a comparison of the magnitudes produced from different stellar population models by calculating the luminosity evolution of a 1 $M$\textsubscript{$\sun$} stellar particle. To ensure consistent comparison, we assume a solar metallicity with a Salpeter IMF and calculate dust-free magnitudes in the SDSS r-band \citep{sdss_photometry}. We select and compare all models that feature this IMF as an option, which include BC03, M05, CB07, C09, and P2. We then use Equation \ref{eq:mag_calc} to compute magnitudes at each age given the SEDs from the model files and the SDSS r-band filter response curve. We convert these magnitudes to a luminosity and present the results in Figure \ref{fig:model_comp}. For the purpose of this demonstration, we calculate magnitudes in the observer's frame assuming a cosmology of $\Omega_m = 0.307$, $\Omega_\Lambda = 0.693$, $h = 0.678$.

\subsection{Set of filters}
\label{sec:filters}

\pymgal\ requires a filter in order to calculate magnitudes. The program contains many filters across different systems, instruments, and surveys. The transmission through the filter is defined by a response curve $R(\nu)$ or $R(\lambda)$, where $\nu$ and $\lambda$ are the frequency and wavelength, respectively. This response curve is read and is used to calculate brightness as described by Equation \ref{eq:mag_calc}.

Many of these curves are obtained from the Flexible Stellar Population Synthesis (FSPS) software by \cite{2010ascl.soft10043C}. These include the Johnson-Cousins system described in \cite{bessell_1990}, the Bessell filters from \cite{bessell_brett_1988}, and the Strömgren filters from \cite{bessell_2011}. It also includes the Steidel set from \cite{steidel_2003}, the Buser filter taken from \cite{bc03}, and the idealised bandpass system.

Filters from various telescopes and surveys are also included. This includes the Two Micron All-Sky Survey \cite[2MASS;][]{2mass}, the Sloan Digital Sky Survey \cite[SDSS;][]{sdss}, the Galaxy Evolution Explorer \cite[GALEX;][]{galex}, the Dark Energy Camera (DECam) from the Dark Energy Survey \cite[DES;][]{des}, the Wide Field Camera (WFCAM) from the United Kingdom Infrared Telescope \cite[UKIRT;][]{ukirt}, the Wide-field Infrared Survey Explorer \cite[WISE;][]{wise}, the Swift Ultraviolet and Optical Telescope \cite[UVOT;][]{uvot}, the Submillimetre Common-User Bolometer Array \cite[SCUBA;][]{scuba} from the James Clerk Maxwell Telescope (JCMT), and the Infrared Astronomical Satellite \cite[IRAS;][]{iras}. It also contains the National Optical Astronomy Observatory's (NOAO) Extremely Wide Field Infrared Imager \cite[NEWFIRM;][]{newfirm}, the Visible and Infrared Survey Telescope for Astronomy \cite[VISTA;][]{vista} and its VISTA Infrared Camera (VIRCAM), the Suprime-Cam from the Subaru Telescope \citep{subaru_suprimecam}, the Panoramic Survey Telescope and Rapid Response System 1 \cite[Pan-STARRS1;][]{panstarrs1}, the Vera C. Rubin Observatory's Large Synoptic Survey Telescope \cite[LSST;][]{lsst}, as well as the Euclid \citep{euclid} and Roman \citep{roman} space telescopes.

Some of these telescopes or surveys contain multiple instruments. The Hubble Space Telescope (HST) is an example of this, as it features the Wide Field and Planetary Camera 2 (WFPC2), the Advanced Camera for Surveys (ACS), the Near Infrared Camera and Multi-Object Spectrometer (NICMOS), and the Wide Field Camera 3 (WFC3), which contains channels for ultraviolet and visible light (UVIS) as well as infrared (IR) \cite[e.g.][]{hst_wfpc2,hst_acs,hst_nicmos,hst_wfc3}. This list also encompasses: the Near Infrared Camera (NIRCam) from the James Webb Space Telescope \cite[JWST;][]{jwst}; the Infrared Array Camera \cite[IRAC;][]{spitzer_irac} and Multiband Imaging Photometer for Spitzer \cite[MIPS;][]{spitzer_mips} from the Spitzer Space Telescope; the Infrared Spectrometer And Array Camera \cite[ISAAC;][]{isaac} and Focal Reducer and low dispersion Spectrograph \cite[FORS;][]{fors} from the Very Large Telescope (VLT); the Photoconductor Array Camera and Spectrometer \cite[PACS;][]{pacs} and the Spectral and Photometric Imaging Receiver \cite[SPIRE;][]{spire} from the Herschel Space Observatory; and the Canada-France-Hawaii 12K camera \cite[CFH12K;][]{cfh12k} and MegaCam \citep{megacam} from the Canada-France-Hawaii Telescope (CFHT).

In addition to those provided by FSPS, we also add filters from the Chinese Space Station Telescope \cite[CSST;][]{csst}, and bands from the Beijing-Arizona Sky Survey (BASS) and the Mayall z-band Legacy Survey (MzLS), which are used in the Dark Energy Spectroscopic Instrument \cite[DESI;][]{desi} survey.

The full set of filters can be found in Table \ref{tab:filters}, with an updated version being maintained at the documentation website. In addition to the currently existing set, custom filters may also be added by the user.

\subsection{Output units}
\label{sec:outputs} 

\pymgal\ provides the option to convert between various output units to facilitate the comparison with observed data. We divide this into three main classes: luminosity, flux or flux density, and magnitude. A summary of available units can be found in Table \ref{tab:outputs}.

\begin{table}
\centering
\begin{tabular}{|l|c|l|}
\toprule
\textbf{Class}  &   \textbf{Subclass}   & \textbf{Units}                   \\
\midrule
Luminosity               &      \hspace{-1em} L                     & erg s\textsuperscript{-1} \\
                         &     L$_{\sun}$                      & L$_{\sun}$ \\
\midrule
Flux/Flux Density        &  F\phantom{$_\nu$}   & erg s\textsuperscript{-1} cm\textsuperscript{-2}   \\
                         & F$_{\nu}$  & erg s\textsuperscript{-1} cm\textsuperscript{-2}  Hz\textsuperscript{-1} \\
                         &  F$_{\lambda}$  & erg s\textsuperscript{-1} cm\textsuperscript{-2}  \AA\textsuperscript{-1} \\
                         & Jy                  & Jansky \\
\midrule
Magnitude                & AB & - \\
                         & Vega   & - \\
                         & Solar   & - \\
\bottomrule
\end{tabular}
\caption{Available output units. Note that magnitudes in any system can be set to either absolute or apparent.}
\label{tab:outputs}
\end{table}

The first option is luminosity. This may be useful if the user wishes to study the intrinsic energy output of the object, with no dependence on the distance at which the projection is held. This can provide a direct measurement of the total energy output of stars within the simulation. Available units in this category are  $\text{erg s}\textsuperscript{-1}$ or solar luminosity L$_{\sun}$. 

The second category is flux and spectral flux density. We denote the flux by the symbol $F$. Since flux is the luminosity per unit area, it is related to luminosity given $F = \frac{L}{4\pi d^2}$ and is expressed in units of $\text{erg s}\textsuperscript{-1}\text{cm}\textsuperscript{-2}$.

An alternative to flux is the spectral flux density, which can measure either the flux per unit wavelength of the flux per unit frequency. We denote the former $F_{\lambda}$ and the latter $F_{\nu}$. Another common unit for flux density is the jansky, typically denoted Jy. The jansky is equivalent to $F_{\nu}$ within a multiplicative factor of $10\textsuperscript{-23}$. $F_\nu$ and $F_{\lambda}$ at a given wavelength are related by $ F_{\lambda} = \frac{c}{\lambda^2} F_{\nu}$.

The final category is the magnitude system. These magnitude systems are logarithmically scaled and defined by some choice of reference brightness. The brightness of an object is then compared with that of the reference and can either be apparent or absolute. 

The absolute magnitude is defined as the brightness that an object would have if it were placed 10 pc away from the observer. By doing so, it eliminates the effect of distance when calculating magnitudes. For apparent magnitude, the distance of the object does affect its value. The apparent magnitude can be obtained by calculating the absolute magnitude and adding a distance term as shown in Equation \ref{eq:app_vs_abs_mag}.

\begin{equation}\label{eq:app_vs_abs_mag}
    m_{\text{apparent}} = m_{\text{absolute}} + 5\log\left(\frac{d}{10 \text{ pc}}\right)
\end{equation}

We include three common magnitude systems. The default is AB magnitude, which uses a reference point of 3631 Jy that is independent of filter choice. The Vega magnitude system is also supported. In this system, the zero point is equal to the brightness of the star Vega in the given filter. The third is the solar magnitude system, which uses the magnitude of the Sun in a given filter to serve as a reference.

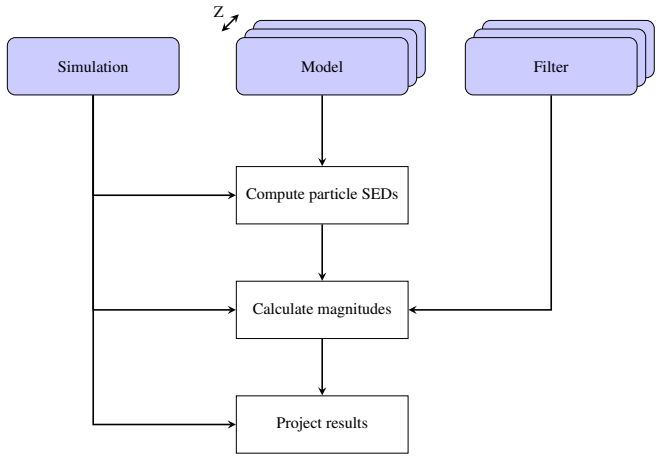
\begin{figure}
    \centering
    \resizebox{\columnwidth}{!}{ 
       \begin{tikzpicture}[node distance=2cm]

    \tikzstyle{input} = [rectangle, minimum width=3cm, minimum height=1cm, text centered, draw=black, fill=blue!20, rounded corners] 
    \tikzstyle{process} = [rectangle, minimum width=3cm, minimum height=1cm, text centered, draw=black, fill=white]
    \tikzstyle{arrow} = [thick,->,>=stealth]
    \tikzstyle{doublearrow} = [thick,<->,>=stealth]

    \node (snapshot) [input] {Simulation};

    \node (sed3) [input, right of=snapshot, xshift=2.3cm, yshift=0.3cm] {}; 
    \node (sed2) [input, right of=snapshot, xshift=2.15cm, yshift=0.15cm] {}; 
    \node (sed) [input, right of=snapshot, xshift=2cm] {Model}; 
    
    \node (filter3) [input, right of=sed, xshift=2.3cm, yshift=0.3cm] {}; 
    \node (filter2) [input, right of=sed, xshift=2.15cm, yshift=0.15cm] {}; 
    \node (filter) [input, right of=sed, xshift=2cm] {Filter};

    \node (combine) [process, below of=sed, yshift=-0.25cm] {Compute particle SEDs};
    \node (calculate) [process, below of=combine] {Calculate magnitudes};

    \node (stop) [process, below of=calculate] {Project results};

    \draw [arrow] (snapshot) |- (combine);
    \draw [arrow] (snapshot) |- (calculate);
    \draw [arrow] (sed) -- (combine);
    \draw [arrow] (filter) |- (calculate);
    \draw [arrow] (combine) -- (calculate);
    \draw [arrow] (calculate) -- (stop);
    \draw [arrow] (snapshot) |- (stop);

    \draw [doublearrow] ([xshift=-0.25cm, yshift=0.1cm] sed3.north west) -- 
        node[midway, above, xshift=-0.2cm] {Z} ([xshift=-0.25cm, yshift=0.1cm] sed.north west);

    \end{tikzpicture}}

    \caption{A flowchart demonstrating the way \pymgal\ processes its input. This example shows the process for models containing multiple metallicities (denoted Z) with magnitudes being calculated over several filters. However, the program can also be run with a single model file and a single filter. }
    \label{fig:flowchart}
\end{figure}

Once the magnitudes have been calculated, the final step is to project the particles to two dimensions. Though we discuss the projection process in greater detail in Section \ref{sec:projections}, we close this section by summarising the way in which data products are read and handled by the program. A flowchart demonstrating these processes can be seen in Figure \ref{fig:flowchart}.


\begin{figure*}
	\includegraphics[width=\textwidth]{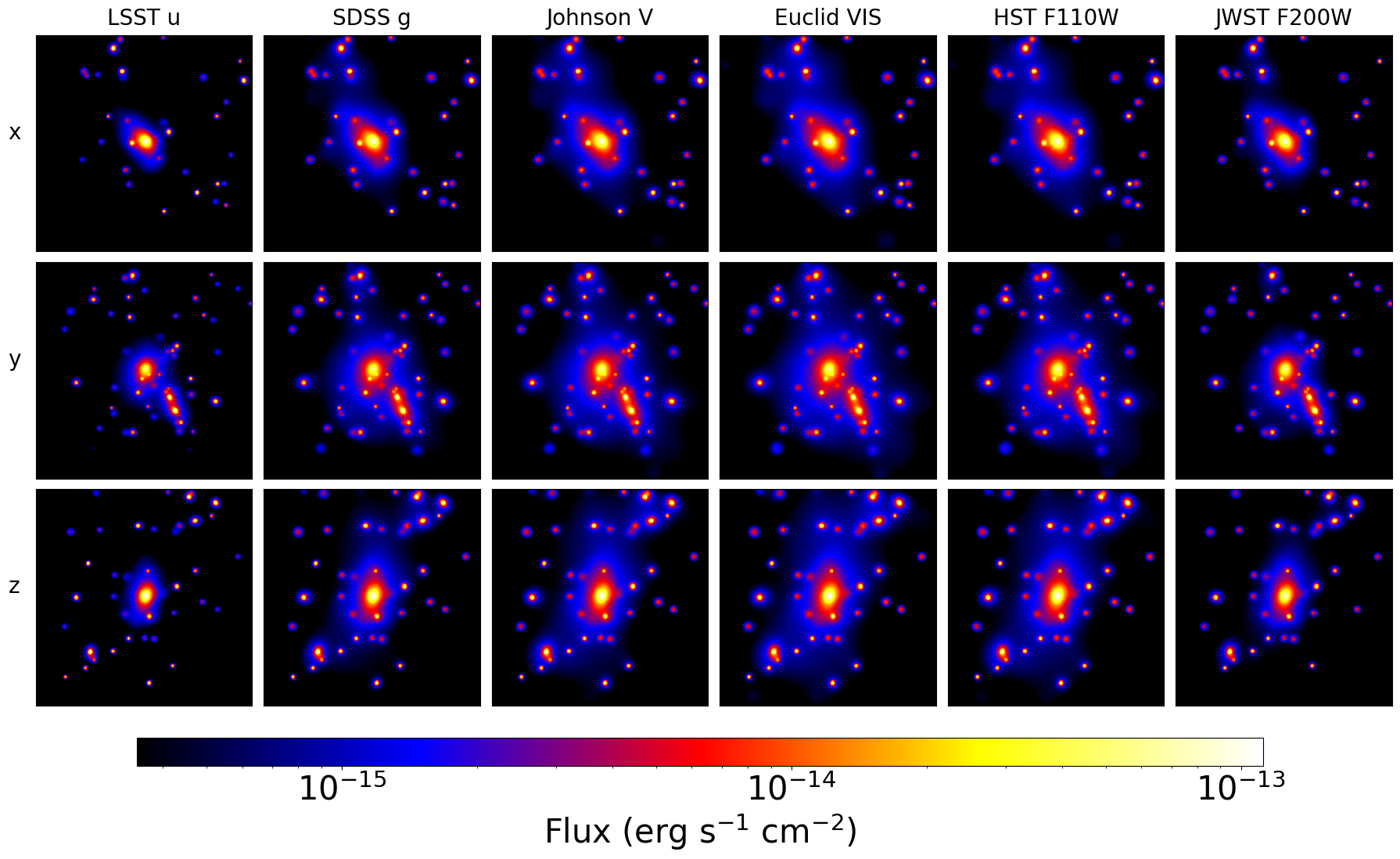}
    \caption{A sample image of a galaxy cluster along different axes and different filters. The three rows show projections along the three principal axes, while the columns indicate different filters from different surveys and instruments. From left to right, the pivot wavelength of the filter becomes larger. Note that the field of view and angular resolution are fixed and identical across all images.}
    \label{fig:demo}
\end{figure*}

\section{Projections}\label{sec:projections}

In this section, we will perform a variety of different tests to demonstrate the capabilities and features of this software. To avoid unnecessary repetition of the parameters of these tests, we define a default configuration. For the simulation data, we use a simulated galaxy cluster from \textsc{The Three Hundred} project \citep{the300}. We select the most massive cluster from the first resimulation region of the GIZMO-SIMBA run \citep{Cui2022}. We select the 118\textsuperscript{th} snapshot, which corresponds to a redshift of $z \approx  0.25$. We set the centre of these projections to be the cluster's halo centre as defined by the Amiga Halo Finder \citep{ahf}. As a default model, we select a dust-free BC03 model with a Chabrier IMF, and we set the default filter to be the SDSS r-band. We set the smoothing length to be the distance to a particle's 100$\textsuperscript{th}$ nearest neighbour and choose flux in $\text{erg } \text{s}\textsuperscript{-1} \text{cm}\textsuperscript{-2}$ to be the default output units. We produce 256x256 pixel images representing a physical side length of 1.5 Mpc.

\subsection{Projecting along an axis}\label{sec:proj_overview}

Once the brightness of particles in the appropriate units is known, mock observations can be generated by projecting the three-dimensional region to two dimensions. To do this, an axis of projection must be specified. The principal axes $x$, $y$, and $z$ are intuitive choices that may be selected. Additionally, \pymgal\ supports arbitrary projection directions, allowing for the simulation data to be observed through any possible angle. 

With an axis specified, the positions of particles are projected to two dimensions. Mock observations are produced by binning these particle brightness values to a 2D histogram. There are three parameters that will affect the way in which simulated particles will be binned. The first of these is the number of bins in the histogram, which is equivalent to the number of pixels in the output image. The other two are the angular resolution and the distance at which the object is held. Further details on the pixel size and the angular resolution of the image can be found in Section \ref{sec:resolution}, while details on the distance of the observation can be found in Section \ref{sec:zobs}.

Each particle is then assigned to a bin in the histogram given its coordinates from the simulation file. The brightness of the particle is then added to the histogram bin, and this process is repeated for all particles.

We provide a first example of these projections in Figure \ref{fig:demo}. We demonstrate a projection of the simulated data along the three principal axes and use filter response curves from multiple different projects and instruments. To show the brightness of the simulated galaxy cluster across multiple wavelengths, we select filters ranging from ultraviolet to infrared. This decreasing frequency trend can be seen from left to right in the figure's columns. Furthermore, to display the ability of the software to mimic different surveys, each frequency uses a filter from a different system, survey, or instrument. From left to right, we show the u-band from LSST, the g-band from SDSS, the V-band from the Johnson system, the VIS band from Euclid, the F110W band from HST's Wide Field Camera 3, and the F200W band from JWST's Near Infrared Camera.

\subsection{Dimensions and angular resolution}
\label{sec:resolution}

The user may select a custom angular resolution and pixel dimension for the output image. The angular resolution, represented by a pixel scale, defines the smallest angular separation between two objects that can be distinguished in the resulting mock observation. A finer angular resolution results in a finer grid of pixels when binning the particles to a 2D histogram. The angular resolution may be set to match that of the instrument they wish to mimic. If an angular resolution is not specified, it will be automatically calculated to encompass all particles within the selected region.

Another parameter is the number of pixels in the image. For a fixed angular resolution, increasing the number of pixels will increase the field of view of the observation. As a result, the total angular size of the image depends on both the pixel scale and the number of pixels. In addition to this, the distance of the simulated region to the frame of observation described in Section \ref{sec:zobs} will also affect the physical appearance of the mock observation. It is therefore possible that the area covered by the output image does not correspond to the exact user-specified region defined in Section \ref{sec:seds}. If the field of view is smaller than this region, the mock observation will only cover a subset of it. If the field of view is larger, pixels in regions without particles are set to zero.

\subsection{Smoothing}
\label{sec:smoothing} 

\pymgal\ performs Gaussian smoothing on particles using the $k$-nearest neighbours (kNN) approach. For each stellar particle, the distance to its $k$\textsuperscript{th} nearest neighbour in the simulation box is calculated and converted to a pixel value. This distance is then used as one standard deviation of a 2D Gaussian distribution, which is then multiplied by the value of the pixel. Because the integral of the normal distribution is equal to one, this ensures that the total quantity within the pixel, such as luminosity, flux, mass, or other, is conserved. The user may define the value of $k$, which will affect how smooth the particles become. The user may also opt to forgo smoothing altogether. In this case, the particles are binned into 2D projections and are left as individual pixel values.

This smoothing method aims to ensure that the light emitted by each particle behaves in a manner more consistent with physical expectations. Rather than being confined to a single point, light is spread over some area, with its intensity decreasing as it travels further away from its source. This also helps remove sharp discontinuities between adjacent pixels resulting from binning particles into a discrete pixel space. Moreover, by using adaptive smoothing scales based on local particle density, this method preserves local features in dense regions while mitigating noise or artifacts in sparse regions.

For mass, age, and metallicity maps, it may not make physical sense to spread the particles as though they are light, as these properties do not travel in the same way as light. In this case, there exists another parameter. The maximum size of kernels for these quantities can be limited by the gravitational softening length. This softening length is then converted to a pixel value during projection. This pixel value is set to be the maximum standard deviation for the Gaussian kernel used to spread mass, age, and metallicity. Further details on these quantities are presented in Section \ref{sec:maps}.

The user may also apply smoothing by specifying an array representing the point spread function (PSF) that may be customised to match the characteristics of the instrument being simulated. This PSF will then be used to convolve the image and spread the intensity of each pixel. By incorporating an appropriate PSF, users can more precisely simulate how the instrument interacts with light, allowing for a more faithful reproduction of how the observed object would appear in actual observations.

\begin{figure}
    \hspace{1.cm}\includegraphics[width=0.9\columnwidth]{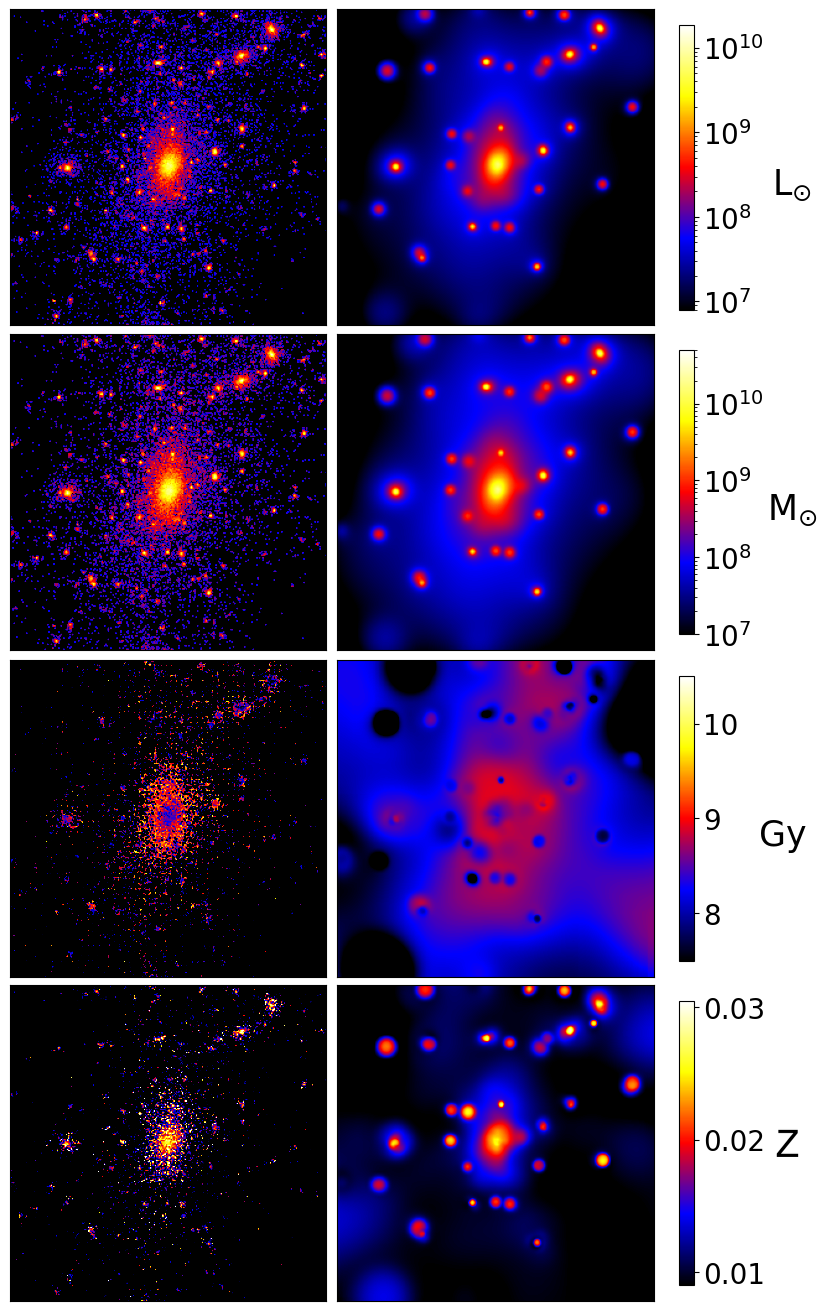}
    \caption{A sample projection of luminosity, mass, age, and metallicity. The first row shows the luminosity in units of $L_{\sun}$, while the second row shows the mass in $M_{\sun}$. The third and fourth rows show the age in Gy and the metallicity, respectively. The left column shows an image without smoothing, and the right column contains the same image smoothed with a length equal to the distance between a particle and its 100$\textsuperscript{th}$ nearest neighbour. Restrictions on the smoothing length are not applied to the mass, age, and metallicity maps.}
    \label{fig:mass_met_age}
\end{figure}

\subsection{Mapping mass, age, and metallicity}
\label{sec:maps} 

In addition to projections of luminosity, magnitudes, or flux through a given filter, \pymgal\ also provides the option to output maps of mass, age, and metallicity. This is an optional output that is calculated after the positions of the particles have been calculated in a 2D projection. In other words, each unique projection has a corresponding map for mass, age, and metallicity.

To calculate the mass inside a pixel, the masses of all particles within the pixel are added together. In the case of age and metallicity maps, values are mass-weighted. Within a pixel, the age or metallicity of a particle is multiplied by its mass. All of these values are then summed together and then divided by the total mass. These relationships can be described mathematically using Equations \ref{eq:mass_map}, \ref{eq:age_map}, and \ref{eq:met_map}, where $m$, $A$, and $Z$ represent mass, age, and metallicity respectively, and $N$ is the number of particles assigned to a pixel.

\begin{align}
    m_{\text{px}} &= \sum_{i=1}^N m_i \label{eq:mass_map} \\
    A_{\text{px}} &=   \frac{ \sum_{i=1}^N A_i m_i}{ \sum_{i=1}^N m_i} \label{eq:age_map} \\
    Z_{\text{px}} &=   \frac{ \sum_{i=1}^N Z_i m_i}{ \sum_{i=1}^N m_i} \label{eq:met_map}
\end{align}

These maps can also be smoothed using the techniques described in Section \ref{sec:smoothing}. For mass maps, the mass pixels are replaced by Gaussian kernels and are smoothed similar to light. For age maps, the product of the age and the mass $A_i m_i$ is computed and is then replaced by a Gaussian kernel. It is then divided by the Gaussian kernel of $m_i$, which has the same size as its numerator. An analogous process is used for metallicity maps. The smoothing of these maps may also be limited by a gravitational softening length as described in Section \ref{sec:smoothing}. An example of these maps and their smoothing is shown in Figure \ref{fig:mass_met_age}.

\subsection{Generating spectra}
\label{sec:spectra}

\begin{figure}
    \centering
    \includegraphics[width=0.8\columnwidth]{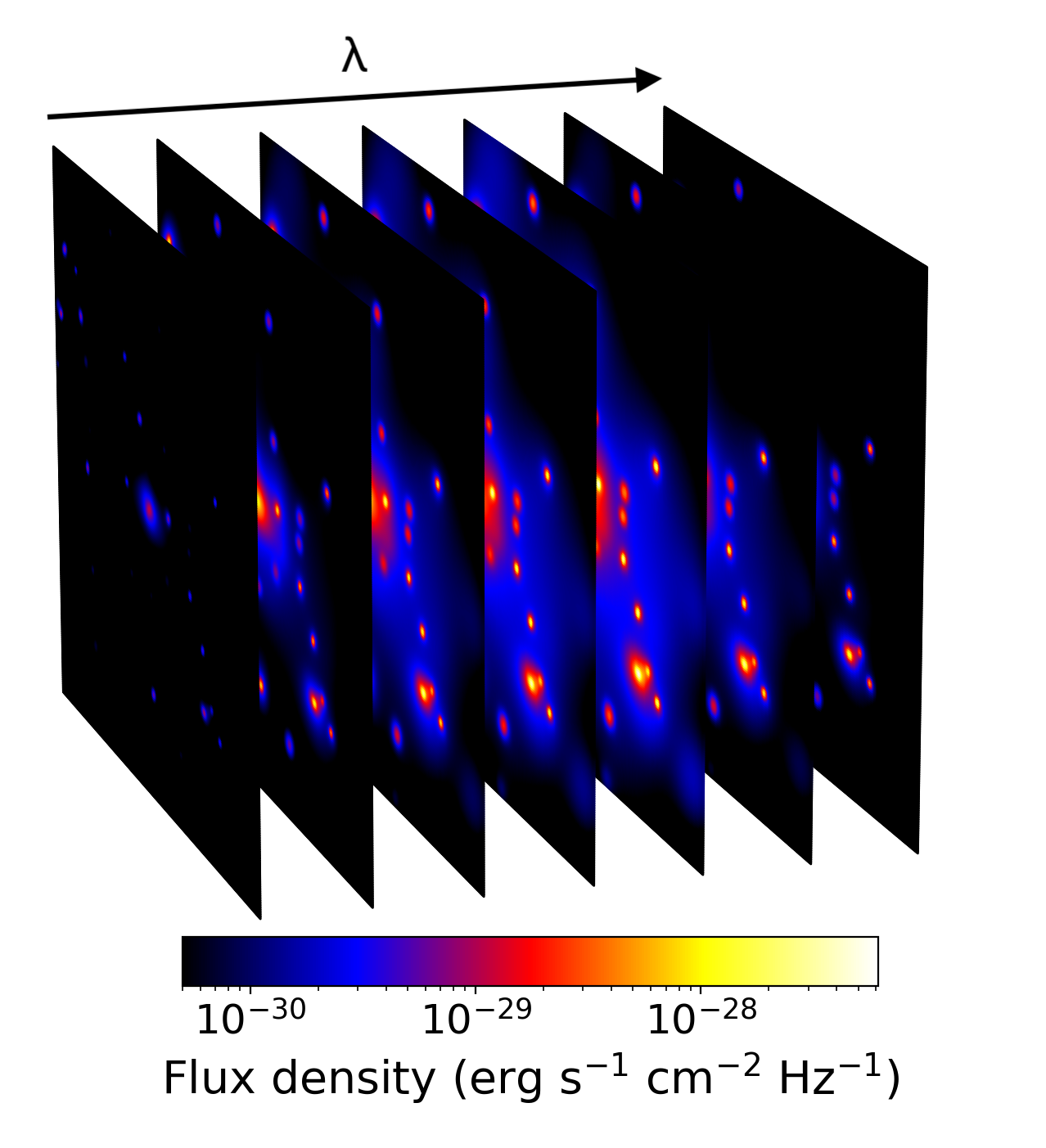}
    \caption{A visualisation of a spectral data cube created by \pymgal. A sample of the full spectral range is shown, with each slice representing the galaxy cluster at a specific wavelength. From left to right, the slices represent wavelengths of approximately 300, 450, 650, 900, 1500, 3000, and 5000 nanometres. Note that this is the intrinsic spectrum and does not represent an observation through a particular filter.}
    \label{fig:sed}
\end{figure}

Another feature supported by \pymgal\ is the ability to output a data cube containing the intrinsic spectrum of the object being observed. The resulting data product is a series of two-dimensional slices stacked along a third spectral dimension. This allows the user to view the object along the entire wavelength range covered by the SEDs of the stellar particles within the simulation. For a given wavelength, \pymgal\ reads the brightness of each particle at that wavelength based on its SED and generates a mock observation by binning the particles in the same manner as shown in Section \ref{sec:proj_overview}. This process is then repeated for each wavelength. A visualisation of this feature is shown in Figure \ref{fig:sed}.

The spacing between wavelengths is defined by the spectral resolution. The maximum possible spectral resolution is that of the SEDs of the stellar particles, though this resolution may be modified. For instance, the spectral resolution may be decreased to match the user's research goals or reduce the memory footprint. Doing so will increase the wavelength difference between neighbouring observations and will therefore reduce the number of wavelengths in the resulting output. The user may also specify a list of wavelengths, in which case only those specified will be selected for the output.

\subsection{Observed redshift}
\label{sec:zobs} 

The observed redshift is a parameter that can be modified to move the object nearer or farther away from the observational frame. Specifying this distance will modify the apparent distance of the mock observations, spectra, and maps of mass, age, and metallicity. This feature may be used for a variety of exercises, such as rescaling for apparent brightness and angular size.

It is important to note that placing objects at distances that differ from their simulation redshift should be treated purely as an observational exercise. This feature does not affect the physical properties or evolutionary traits of the simulation and should not be used in an attempt to realistically simulate an object at higher or lower redshift than that of the simulation snapshot. The user is advised to keep the same redshift as the simulation when producing mock observations that attempt to replicate real systems.

\begin{figure}
	\includegraphics[width=\columnwidth]{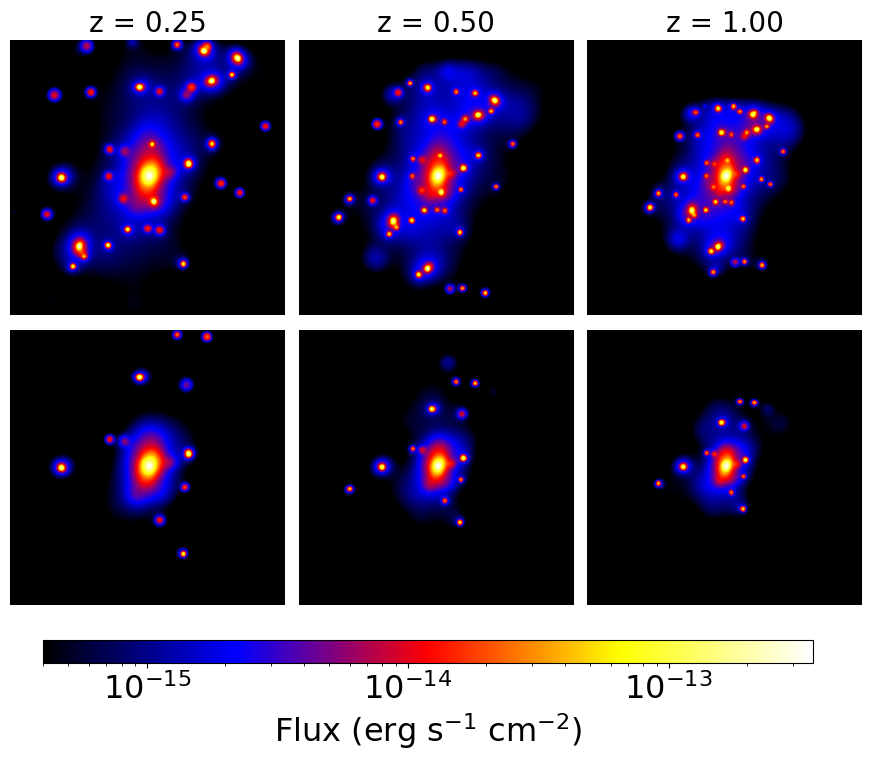}
    \caption{A demonstration of selecting different redshift values and thickness cuts for observation. The cluster is shown at different distances for each column, with $z=0.25$ on the left, $z=0.50$ in the centre, and $z=1.00$ on the right. In the top row, the entire thickness of the region is selected, while in the bottom row, a cut of 100 kpc is taken in the projection direction.}
    \label{fig:z_obs_thick}
\end{figure}

We demonstrate this effect in the columns of Figure \ref{fig:z_obs_thick}. We show a projection of the same simulation region in units of flux for three different redshift values. The rows of this figure display different choices of thicknesses along the axis of projection, which is described in Section \ref{sec:thickness}.

\subsection{Thickness}
\label{sec:thickness} 

The user can optionally select to project only a slice of the data along the axis of projection. By default, all particles in the specified field of view are shown in the resulting maps. However, a cut can be made along the depth axis to limit its thickness. If a thickness is specified, only particles falling within the projection centre plus or minus the depth will be displayed in the output image. Any particle falling beyond this range will not be shown. This feature can be used for different research purposes, such as to help focus on a specific region or omit foreground and background objects.

This effect is illustrated in the rows of Figure \ref{fig:z_obs_thick}. In the top row, we project the entire simulated region for different redshift values as described in Section \ref{sec:zobs}, with no restriction placed on thickness. In the bottom row, we restrict the thickness to 100 kpc. While the two settings produce similar images, we find that some galaxies are omitted when the thickness is limited.

\subsection{Adding noise}
\label{noise}

Noise may also be added to the observations to add realism and make them align more closely with real observations. \pymgal\ features the option to add adjustable levels of Gaussian noise that can be tailored to match the observational scenario being modelled. This level can be modified on a continuous scale and is defined as a limiting AB magnitude at a chosen signal-to-noise ratio (SNR) within a chosen circular aperture.

To model this, \pymgal\ begins by converting the limiting AB magnitude into the user's selected output units. It then scales this flux value based on the SNR and the area of the aperture. We can define this mathematically as follows, where we take $m_{lim} $ to be the limiting AB magnitude, $g$ to be a function that maps this magnitude to the appropriate output units, SNR to be the signal-to-noise ratio, and $r_{\text{ap}}$ to be the radius of the aperture converted into a pixel value. 

\begin{equation}
    \sigma(m_{lim}, \text{SNR}, r_\text{ap}) = \frac{g(m_{lim})}{ \text{SNR } \sqrt{\pi} r_\text{ap}}
\end{equation}

This value is then used as the standard deviation for a Gaussian distribution used to combine the image with noise. Should the selected output unit be a magnitude, where smaller values correspond to higher brightness, the values are first converted into flux, combined with noise, and then converted back to magnitude.

\begin{figure}
        \centering
	\includegraphics[width=\columnwidth]{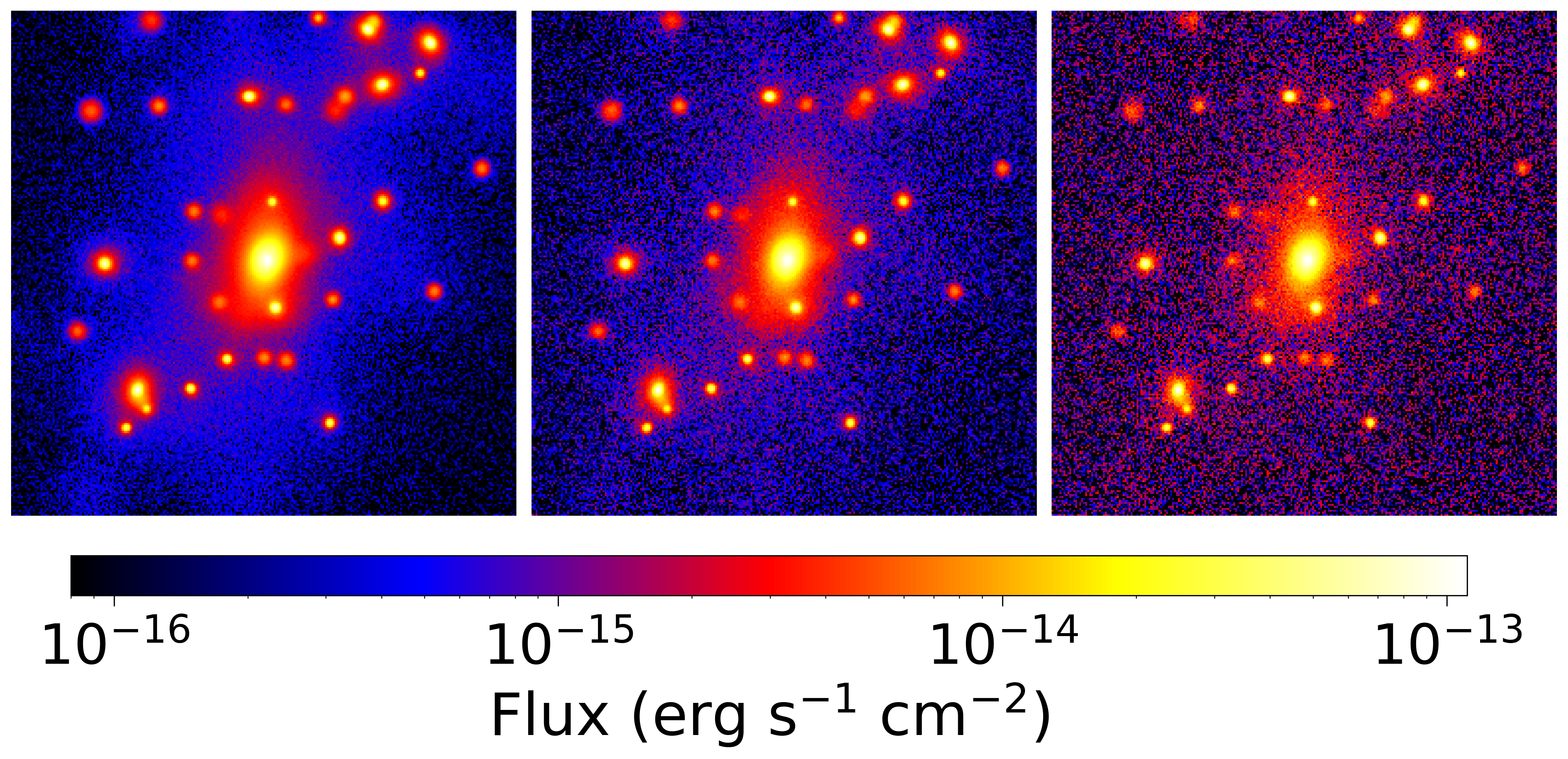}
    \caption{A demonstration of different noise levels added to the image. From left to right, a limiting AB magnitude of 27, 26, and 25 is selected with a constant signal-to-noise ratio of 5 within a circular aperture of radius 0.25 arcseconds. }
    \label{fig:noise}
\end{figure}

\begin{figure*}
    \centering
    \includegraphics[width=\textwidth]{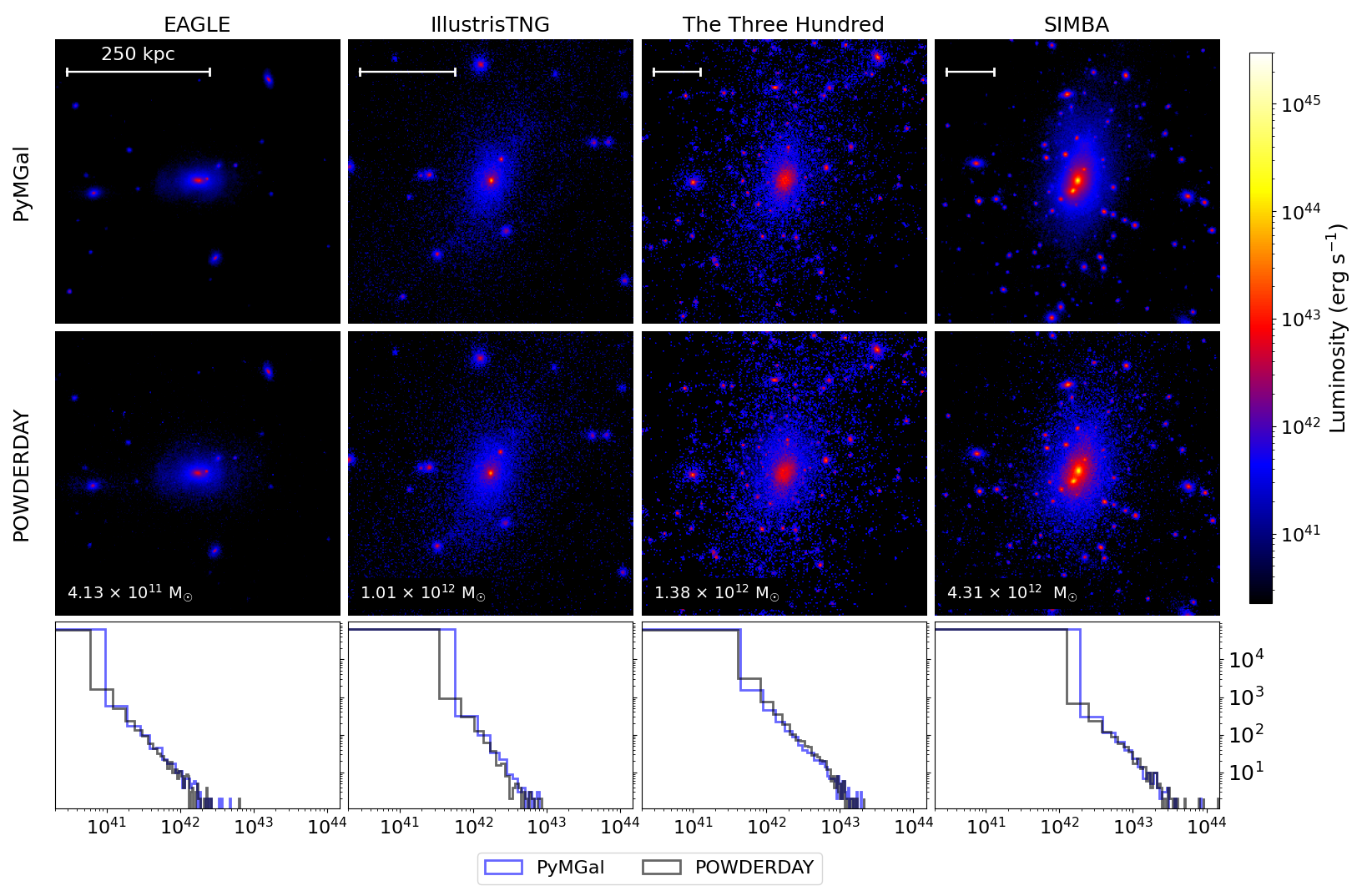}
    \caption{A comparison of \pymgal's outputs in the first row using the radiative transfer package \textsc{powderday} in the second row as a benchmark. The third row displays the luminosity distribution in pixel counts for both packages. From left to right, we display the most massive halo across $z\approx 0.25$ snapshots from the EAGLE, IllustrisTNG, \textsc{The Three Hundred}, and SIMBA hydrodynamical simulations, ordered by increasing stellar mass within a 50 kpc radius of the halo centre. The stellar mass within this region is indicated in the bottom-left corner of each subplot in the second row, while a 250 kpc ruler is included in the top-left corner of each subplot in the first row for size reference.}
    \label{fig:benchmark_comp}
\end{figure*}

We demonstrate this effect in Figure \ref{fig:noise}, where three gradually increasing noise levels are presented from left to right. We select a constant signal-to-noise ratio of 5 and a constant aperture radius of 0.25 arcseconds. In each case, we amplify the noise level by one AB magnitude.

Users requiring more detailed noise characteristics are encouraged to obtain noise images directly from the instrument and filter being modelled. To more accurately replicate observational conditions, mock images may be embedded into noise frames in order to incorporate elements such as foreground objects and instrumental artifacts.

\section{Validation} \label{sec:validation}

In this section, we demonstrate the fidelity of \pymgal's outputs by comparing them with an established benchmark. We use \textsc{powderday} as this benchmark, which features a full radiative transfer pipeline with customisable stellar population modelling via FSPS. We perform this comparison across four different simulations of varying box sizes and resolutions. More specifically, we choose volumes from the EAGLE, SIMBA, IllustrisTNG, and \textsc{The Three Hundred} simulations at redshifts of approximately $z\approx 0.25$ to maintain consistency with the previous section. 

For EAGLE, we select the 25\textsuperscript{th} snapshot from the RefL0050N0752 run, which has a box size of 50 Mpc and 2$\times$752\textsuperscript{3} particles (i.e. the same number of dark matter and baryonic components). For SIMBA, we take the 137\textsuperscript{th} snapshot of the flagship run (m100n1024), which has a size of 100 $h$\textsuperscript{-1} Mpc and 2$\times$1024\textsuperscript{3} particles. For IllustrisTNG, we select the 80\textsuperscript{th} snapshot from the intermediate resolution run of TNG300 (TNG300-2), which has a size of approximately 300 Mpc and 2$\times$1250\textsuperscript{3} particles. Finally, for \textsc{The Three Hundred}, we select the same snapshot used throughout Section \ref{sec:proj_overview}, which shows a resimulation region within a 1 $h$\textsuperscript{-1} Gpc box containing 3840\textsuperscript{3} dark matter particles. All above box sizes are reported as comoving distances.

We then run both \pymgal\ and \textsc{powderday} on the most massive halo within each simulation snapshot. We compute the total stellar mass within 50 kpc of the halo centre as an estimate of the central galaxy’s stellar mass. In increasing order of stellar mass, we find values of $4.13\times 10^{11}$ M$_{\odot}$ for EAGLE, $1.01\times 10^{12}$ M$_{\odot}$ for IllustrisTNG, $1.38\times 10^{12}$ M$_{\odot}$ for \textsc{The Three Hundred}, and $4.31\times 10^{12}$ M$_{\odot}$ for SIMBA. We then produce images with physical side lengths of 500 kpc for EAGLE, 750 kpc for IllustrisTNG, and 1.5 Mpc for both \textsc{The Three Hundred} and SIMBA. We project along the z-axis in the SDSS r-band and generate 256x256 pixel images.  To match \textsc{powderday}'s stellar population models, we construct a custom SSP model using FSPS having 30 metallicities ranging from $Z=0.0001$ to $Z=0.05$ that are equally separated on a logarithmic scale. We do not treat effects from nebular emission or black holes, and we select a Chabrier IMF in each case. We generate \textsc{powderday} observations using a dust-to-metals ratio of 0.3, and we generate corresponding \pymgal\ observations using a Charlot \& Fall dust function and a constant-opacity line-of-sight attenuation model matched to the level of \textsc{powderday}. We do not add smoothing, thickness cuts, or noise. The results can be seen in Figure \ref{fig:benchmark_comp}.

We find that the outputs from \pymgal\ are consistent with those from the radiative transfer package and show only minor variations. This can be seen visually by comparing images from \pymgal\ in the first row of Figure \ref{fig:benchmark_comp} with images from \textsc{powderday} in the second row. The luminosity distributions in the third row also exhibit close agreement between the two packages, with both yielding similar luminosity values. While the validation performed in this section assesses the core functionality of the package, users interested in a more detailed assessment of \pymgal's outputs are encouraged to perform additional tests tailored to their scientific uses. These include comparisons at different wavelengths, notably those more affected by dust, statistical analyses of fluxes and colour indices on large samples of sources, and comparisons against observed datasets. Nonetheless, these results demonstrate that \pymgal\ can reliably reproduce synthetic observations while offering significant computational efficiency compared to full radiative transfer methods.

\section{Conclusions}
\label{sec:conclusions}

\pymgal\ provides a quick and flexible way of generating mock observations from cosmological simulations. The functionality of the software can be broadly split into three parts: the modelling of stellar populations, the calculation of magnitudes given a filter, and the projection of the particles to a 2D plane. Each of these parts is detailed in a section of this paper.

The first part, highlighted in Section \ref{sec:modelling}, involves modelling stellar populations and leveraging this to generate SEDs. The software infers the SEDs of stellar particles in simulation snapshot files given an SPS library that can be tailored to fit the user's research goals. While different libraries can be selected, they share a similar structure. Each file consists of a 2D array with ages on one axis and wavelengths or frequencies on the other. The libraries account for the metallicity dependence of these SEDs by dividing the metallicity range into discrete quantities and having one file for each value. \pymgal\ then reads the physical properties of simulated particles and assigns SEDs based on the selected library.  The package supports all SPS libraries included in EzGal and allows the user to implement customised libraries as well.

The second part, shown in Section \ref{sec:filters_and_mags}, describes the way \pymgal\ calculates magnitudes given the SEDs and a filter response curve. The magnitudes computed for different SPS models will vary model by model, but display reasonable agreement when compared. These filters can be selected from a wide range of surveys, systems, and instruments. The package supports several different output units including luminosity, flux, spectral flux density, and magnitude.

Once these magnitudes have been computed, mock observations can be created by projecting the data to two dimensions as described in Section \ref{sec:projections}. This can be done by mapping the physical positions of the particles to a pixel position in a 2D image. \pymgal\ supports different projection features, including the choice of projection axis, image dimensions, angular resolution, and particle smoothing. It also has the ability to create mass, age, and metallicity maps, as well as spectral data cubes for a given mock observation. Finally, the user may also customise the apparent distance of the observation, the thickness in the projection region, and the level of noise.

In summary, \pymgal\ is a versatile tool that bridges cosmological simulations with observational data, allowing researchers to create optical mock observations of galaxies from hydrodynamical simulations based on the stellar population model that best suits their needs. It is adaptable to different simulation formats, output units, SPS models, filter response curves, and more, making it a practical resource for various astrophysical research applications.

\section*{Acknowledgements}

We acknowledge access to the theoretically modelled galaxy cluster data via  \textsc{The Three Hundred}\footnote{\url{https://www.the300-project.org}} collaboration. The simulations used in this paper have been performed in the MareNostrum Supercomputer at the Barcelona Supercomputing Center, thanks to CPU time granted by the Red Española de Supercomputación. As part of The Three Hundred project, this work has received financial support from the European Union’s Horizon 2020 Research and Innovation programme under the Marie Skłodowska-Curie grant agreement number 734374, the LACEGAL project.

PJ acknowledges support from the Centre de recherche en astrophysique du Québec, un regroupement stratégique du FRQNT. He also acknowledges the support of the Natural Sciences and Engineering Research Council of Canada (NSERC), [funding reference number RGPIN-2020-04606]. WC is supported by the Atracci\'{o}n de Talento Contract no. 2020-T1/TIC-19882 granted by the Comunidad de Madrid in Spain, and the Agencia Estatal de Investigación (AEI) for the Consolidación Investigadora Grant CNS2024-154838. He also thanks the Ministerio de Ciencia e Innovación (Spain) for financial support under Project grant PID2021-122603NB-C21 and HORIZON EUROPE Marie Sklodowska-Curie Actions for supporting the LACEGAL-III project with grant number 101086388. This work is supported by the China Manned Space Program with grant no. CMS-CSST-2025-A04.

\section*{Data Availability}

\pymgal\ is registered with the Python Package Index at \url{https://pypi.org/project/pymgal/}, which contains the relevant information, documentation, and links. The documentation can also be directly accessed at \url{https://pymgal.readthedocs.io}. 

This paper uses data from the EAGLE, SIMBA, IllustrisTNG, and \textsc{The Three Hundred} hydrodynamical simulations. All relevant data products used in this work are either publicly available or may be shared upon request to the respective collaboration.



\bibliographystyle{rasti}
\bibliography{rasti_paper.bib} 




\appendix

\section{Table of Available Filters}

We include a table of all filters included in the package upon download. An up-to-date version of this list can be found on the documentation website. The user may also add custom filters if the desired one is not present. 

\begin{table*}
\centering
\begin{tabular}{|l|l|l|}
\hline
\textbf{Group}    & \textbf{Specification}    & \textbf{Filters}           \\
\hline
Johnson-Cousins        &   & U, V, B, R, I                                 \\
\hline
Buser          &  &   B2                                   \\
\hline
Bessell  &   &   \pbox{8cm}{L, L', M}    \\
\hline
Strömgren  &   &  \pbox{8cm}{u, v, b, y}    \\
\hline
Steidel     & & \pbox{8cm}{U$_n$, G, R$_s$, I }    \\
\hline
Idealised bandpass &  &   \pbox{8cm}{1500Å, 2300Å, 2800Å}    \\
\hline 
2MASS       &     & J, H, K$_s$             \\
\hline
SDSS        &   & u, g, r, i, z     \\
\hline
GALEX      &   & \pbox{8cm}{FUV, NUV }    \\
\hline
DES     & DECam &  \pbox{8cm}{g, r, i, z, Y}    \\
\hline
UKIRT &  WFCAM &   \pbox{8cm}{Z, Y, J, H, K}    \\
\hline
WISE    & &  \pbox{8cm}{W1, W2, W3, W4}    \\
\hline
Swift     & UVOT &  \pbox{8cm}{W2, M2, W1}    \\
\hline
JCMT    & SCUBA &  \pbox{8cm}{450WB, 850WB	}    \\
\hline
IRAS    & &  \pbox{8cm}{12$\mu$m, 25$\mu$m, 60$\mu$m, 100$\mu$m}    \\
\hline
NOAO    &  NEWFIRM& \pbox{8cm}{J1, J2, J3, H1, H2, K}    \\
\hline
\pbox{6cm}{VISTA}    & VIRCAM & \pbox{8cm}{Y, J, H, K}    \\
\hline
Subaru     & Suprime-Cam &\pbox{8cm}{B, g, V, r, i, z}    \\
\hline
Pan-STARRS1     & & \pbox{8cm}{g, r, i, z, y}    \\
\hline
LSST  &  &\pbox{8cm}{u, g, r, i, z, y}    \\
\hline
Euclid   &    &\pbox{8cm}{VIS, Y, J, H}    \\
\hline
Roman    &  &\pbox{8cm}{F062, F087, F106, F129, F158, F184}    \\
\hline
HST    &  WFPC2 &  \pbox{10cm}{F255W, F300W, F336W, F439W, F450W, F555W, F606W, F814W, F850LP }    \\ \cmidrule{2-3}
        &  ACS  & \pbox{10cm}{F435W, F475W, F555W, F606W, F625W, F775W, F814W, F850LP }    \\ \cmidrule{2-3}
   &   WFC3 UVIS    & \pbox{8cm}{F218W, F225W, F275W, F336W, F390W, F438W, F475W, F555W, F606W, F775W, F814W, F850LP }    \\ \cmidrule{2-3}
       & WFC3 IR   & \pbox{6cm}{F098M, F105M, F110M, F125M, F140M, F160M }    \\ \cmidrule{2-3}
      & NICMOS & \pbox{6cm}{F110W, F160W }    \\
\hline
JWST    & NIRCam &  \pbox{9cm}{F070W, F090W, F115W, F150W, F200W, F277W, F356W, F444W}    \\
\hline
Spitzer     &  IRAC  & \pbox{6cm}{3.6$\mu$m, 4.5$\mu$m, 5.8$\mu$m, 8.0$\mu$m }    \\ \cmidrule{2-3}

            &  MIPS &  \pbox{6cm}{24$\mu$m, 70$\mu$m, 160$\mu$m}    \\
\hline
VLT       &  ISAAC &  \pbox{8cm}{K$_s$ }    \\ \cmidrule{2-3}
                                &  FORS &  \pbox{8cm}{V, R }    \\ 
\hline
CFHT        &  CFH12K  & B, R, I                 \\ \cmidrule{2-3}
        &  MegaCam & \pbox{8cm}{u, g, r, i, z}    \\
\hline
Herschel     & PACS & \pbox{8cm}{70$\mu$m, 100$\mu$m, 160$\mu$m}    \\ \cmidrule{2-3}
              & SPIRE &  \pbox{8cm}{250$\mu$m, 350$\mu$m, 500$\mu$m}    \\ 
\hline
DESI   & BASS  &\pbox{8cm}{g, r}    \\ \cmidrule{2-3}
  & MzLS  &\pbox{8cm}{z}    \\
\hline
CSST  &   &\pbox{8cm}{NUV, u, g, r, i, z, y}    \\
\hline
\end{tabular}
\caption{Available filters. An updated version is maintained at the documentation website \url{https://pymgal.readthedocs.io}.}
\label{tab:filters}
\end{table*}


\bsp	
\label{lastpage}
\end{document}